\documentclass[aps,reprint,showpacs,amsmath,amssymb,prl,superscriptaddress]{revtex4-1}

\usepackage{graphicx}
\usepackage{dcolumn}
\usepackage{bm}
\usepackage{epsf}
\usepackage{amsmath}
\usepackage{amssymb}
\usepackage{color}
\usepackage[colorlinks,pdfpagelabels,pdfstartview = FitV, linkcolor = black,plainpages = true,hypertexnames = false,citecolor = black]{hyperref}
\usepackage{multirow}

\newcommand{\be}{\begin{equation}}
\newcommand{\ee}{\end{equation}}

\newcommand{\la}{\left\langle}
\newcommand{\ra}{\right\rangle}

\newcommand{\etal}{\textit{et al.}}

\newcommand{\ct}[1]{\coth{\left(#1\right)}}
\newcommand{\bla}{bla\\bla\\bla\\bla\\bla}

\newcommand{\PRE}{Phys. Rev. E }
\newcommand{\PRL}{Phys. Rev. Lett. }

\newcommand{\RMP}{Rev. Mod. Phys. }
\newcommand{\NJP}{New. J. Phys. }

\begin{document}

\title{Single ion heat engine with maximum efficiency at maximum power}

\author{O. Abah}
\affiliation{Department of Physics, University of Augsburg, D-86159 Augsburg, Germany}
\author{J. Ro{\ss}nagel}

\affiliation{Institut f\"ur Quantenphysik, Universit\"at Mainz, 55128 Mainz, Germany}

\author{G. Jacob}
\affiliation{Institut f\"ur Quantenphysik, Universit\"at Mainz, 55128 Mainz, Germany}
\author{S. Deffner}
\affiliation{Department of Physics, University of Augsburg, D-86159 Augsburg, Germany}
\affiliation{Department of Chemistry and Biochemistry and
Institute for Physical Sciences and Technology, University of Maryland,
College Park, MD 20742, USA}

\author{F. Schmidt-Kaler}
\affiliation{Institut f\"ur Quantenphysik, Universit\"at Mainz, 55128 Mainz, Germany}

\author{K. Singer}
\affiliation{Institut f\"ur Quantenphysik, Universit\"at Mainz, 55128 Mainz, Germany}

\author{E. Lutz}
\affiliation{Department of Physics, University of Augsburg, D-86159 Augsburg, Germany}
\affiliation{Dahlem Center for Complex Quantum Systems, FU Berlin, D-14195 Berlin, Germany}


\begin{abstract}
We propose an experimental scheme to realize a { nano} heat engine with a single ion.
An Otto cycle may be implemented by confining the ion in a linear Paul trap with  tapered geometry and coupling it to  engineered laser reservoirs. The { quantum} efficiency at maximum power is analytically determined in various regimes. Moreover, Monte Carlo simulations of the engine are performed that demonstrate its feasibility \mbox{and its ability to operate at maximum efficiency  {of   $30\%$ under realistic conditions}}.
\end{abstract}

\maketitle

Miniaturization has lead to the development of increasingly smaller devices. This ongoing size reduction from the macroscale to the nanoscale is  approaching the ultimate limit, given by the atomic nature of matter \cite{cer09}. Prominent macro-devices are heat engines that convert thermal energy into mechanical work, and hence motion \cite{cen01}. A fundamental question is whether these machines can be scaled down to the single particle level, while retaining the same working principles as, for instance, those of a car engine.
It is interesting to note in this context that biological molecular motors are based on completely different mechanisms that exploit the constructive role of thermal fluctuations \cite{kay07,mar09}.
At the nanoscale, quantum properties become important and have thus to be fully { taken into account}.
Quantum heat engines have been the subject of extensive theoretical studies in the last fifty years \cite{sco59,ali79,kos84,gev92,scu02,scu03,hum02,kie04,dil09,gem09}.
However, while classical micro heat engines have been fabricated, using optomechanical \cite{hug02}, microelectromechanical \cite{jac03,wha03, steeneken11}, and colloidal systems \cite{bechinger11}, to date no quantum heat engine has been built. 

\begin{figure}
\includegraphics[width=\columnwidth]{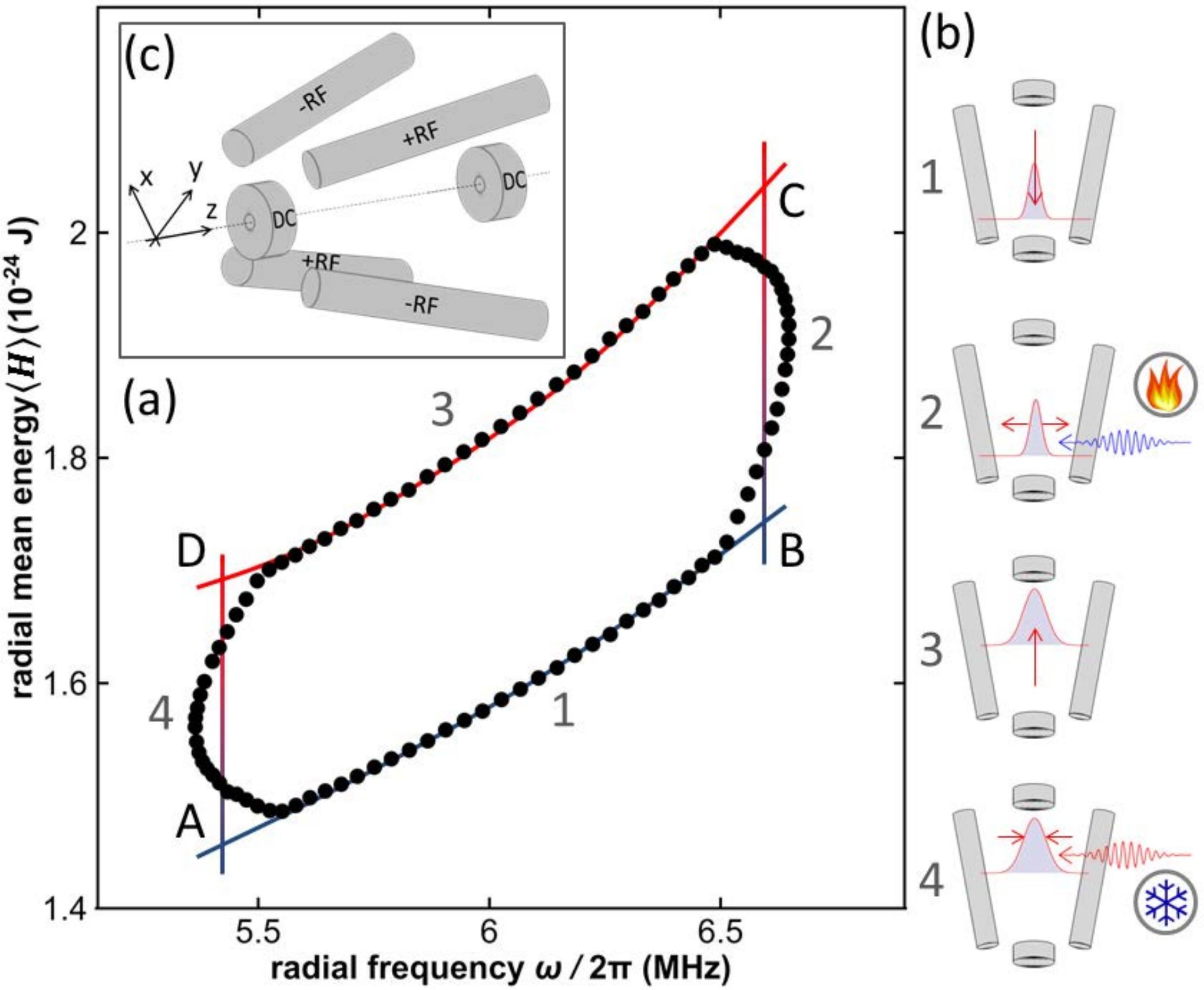}
\caption{(Color online) (a)~Energy-frequency diagram of the Otto cycle for the radial degree of freedom of the ion. The continuous line represents the ideal process, while the dots show the results of the Monte Carlo simulations. (b)~The pictograms illustrate the four individual strokes of the cycle for the radial thermal state. (c)~Geometry of the tapered Paul trap: the rf-electrodes have an angle of $\theta=20^{\circ}$ with the trap axis, the length of the trap is $5\,\mbox{mm}$, the radial distance of the ion to the rf-electrodes is $r_0=1\,\mbox{mm}$.
\label{cycle}}
\end{figure}

In this paper, {we take a step towards that goal by proposing} a single ion heat engine using a  linear Paul trap. Specifically, we present a scheme {which has the potential to implement}  a quantum Otto cycle using currently available state-of-the-art ion-trap technology. Laser-cooled ions in linear Paul traps are quantum systems with remarkable properties \cite{lei03}: they offer an unprecedented degree of preparation and control of their parameters, permit their cooling to the ground state, and allow the coupling to engineered reservoirs \cite{poyatos96}.
For these reasons, they have played a prominent role in the experimental study of quantum computation and information processing applications \cite{blatt08,monz11}.
They are also invaluable tools for the investigation of quantum thermodynamics \cite{hub08}.
The quantum Otto cycle for a harmonic oscillator is a quantum generalization of the common four-stroke car engine and a paradigm for thermodynamic quantum devices \cite{lin03,rez06,qua07}.
It consists of two isentropic processes during which the frequency of the oscillator (the trap frequency) is varied, and of two isochoric processes, that in this situation correspond to a change of temperature at constant frequency, see Fig.~\ref{cycle}(a).
In the present proposal, we simulate the Otto cycle by confining a single ion in a novel trap geometry with an asymmetric electrode configuration (see Fig.~\ref{cycle}(c)) and coupling it alternatingly to two engineered laser reservoirs.
As for all realistic machines, this Otto engine runs in finite time and has thus non-zero power \cite{and84}.
We determine the efficiency at maximum power in the limit of adiabatic and strongly nonadiabatic processes, which we express in terms of the nonadiabaticity parameter introduced by Husimi \cite{hus53}.
We further present semiclassical Monte Carlo simulations, with realistic parameters, that demonstrate the experimental feasibility of such a device. The proposal and the single ion trap design idea have several unique advantages: First, all of the parameters of the engine, in particular the temperatures of the  baths, are tunable over a wide range, in contrast to existing engines. 
As a result, maximum efficiency can be achieved.
Moreover, at low temperatures, the engine {may} operate in the quantum regime, {where the discreteness of the energy spectrum plays an important role}. In addition, the coupling to the laser reservoirs can be either switched on and off externally, or by the intrinsic dynamics of the ion itself. In this latter mode, the heat engine runs autonomously \cite{mahler05}.
{Since  trapped ions are  perfect oscillator models, the results described here may in principle be   extended to  analogous systems, such as micro- and nanomechanical oscillators \cite{OConell10,lin10,kar09,fau12}, offering a broad spectrum of potential applications.}

\paragraph{Quantum Otto cycle.}
We consider a quantum engine whose working medium is a single harmonic oscillator with time-dependent frequency $\omega_t$, changing between $\omega_1$ and $\omega_2$. The engine is alternatingly coupled to two heat baths at inverse temperatures $\beta_i=1/({ k_B} T_i)$ $(i=1,2)$, where ${ k_B}$ is the Boltzmann constant. The Otto cycle consists of four consecutive steps as shown in Fig.~\ref{cycle}(a):\\
(1) \textit{Isentropic compression} A$(\omega_1,\beta_1)\rightarrow B(\omega_2,\beta_1)$: the frequency is varied during time $\tau_1$ while the system is isolated. The evolution is unitary and the von Neumann entropy of the oscillator is thus constant. Note that state B is non-thermal even for slow (adiabatic) processes. \\
(2) \textit{Hot isochore} B$(\omega_2,\beta_1)\rightarrow C(\omega_2,\beta_2)$: the oscillator is weakly coupled to a reservoir at inverse temperature $\beta_2$ at fixed frequency and allowed to relax during time $\tau_2$ to the thermal state C. This equilibration is much shorter than the expansion/compression phases (see  below). \\
(3) \textit{Isentropic expansion} $C(\omega_2,\beta_2)\rightarrow D(\omega_1,\beta_2)$: the frequency is changed back to its initial value during time $\tau_3$. The  isolated oscillator evolves unitarily into the non-thermal state $D$ at constant entropy.\\
(4)  \textit{Cold isochore} $D(\omega_1,\beta_2)\rightarrow A(\omega_1,\beta_1)$:  the system is weakly coupled to a reservoir at inverse temperature $\beta_1>\beta_2$  and quickly relaxes to the initial thermal state A during $\tau_4$. The frequency is again kept constant.

In order to determine the efficiency of the quantum Otto cycle, we need to evaluate work and heat for each of the above steps. During stroke (2) and (4), the frequency is constant, and thus only heat is exchanged with the reservoirs. On the other hand,
during stroke (1) and (3), the system is isolated and only work is performed by modulating the frequency. Since the dynamics is unitary in the latter, the Schr\"odinger equation for the parametric oscillator can be solved exactly and its mean energy can be obtained analytically. The average {quantum} energies {$\la H\ra$} of the oscillator at the four stages of the cycle are
\begin{subequations}
\begin{eqnarray}
\label{1}
\la H\ra_{A}&=&\frac{\hbar\omega_1}{2}\,\ct{\frac{\beta_1\hbar\omega_1}{2}},\\
\la H\ra_{B}&=&\frac{\hbar\omega_2}{2}\,Q^\ast_1\,\ct{\frac{\beta_1\hbar\omega_1}{2}},\\
\la H\ra_{C}&=&\frac{\hbar\omega_2}{2}\,\ct{\frac{\beta_2\hbar\omega_2}{2}},\\
\la H\ra_{D}&=&\frac{\hbar\omega_1}{2}\,Q^\ast_2\,\ct{\frac{\beta_2\hbar\omega_2}{2}},
\end{eqnarray}
\end{subequations}
where we have introduced the two adiabaticity parameters $Q^\ast_1$ and $Q^\ast_2$ \cite{hus53}. They are equal to one for adiabatic (slow) processes and increase with the degree of nonadiabaticity. Their explicit expressions  for any given modulation $\omega_t$, can be found in Refs.~\cite{def08,def10}. {Equations (1a-d) reduce to their classical limits when $\hbar \rightarrow 0$}. The mean work, {denoted by $\la W_1\ra $}, done during the first stroke is
\begin{equation}
\label{2}
\la W_1\ra=\la H\ra_{B}-\la H\ra_{A}
=\left(\frac{\hbar\omega_2}{2}\,Q^\ast_1-\frac{\hbar\omega_1}{2}\right)\ct{\frac{\beta_1\hbar\omega_1}{2}}
\end{equation}
whereas the mean heat {$\la Q_2 \ra $} exchanged with the hot reservoir during the second stroke reads,
\begin{eqnarray}
\label{3}
\la Q_2\ra&=&\la H\ra_C-\la H\ra_B \nonumber\\
&=&\frac{\hbar\omega_2}{2}\left[\ct{\frac{\beta_2\hbar\omega_2}{2}}-Q^\ast_1\,\ct{\frac{\beta_1\hbar\omega_1}{2}}\right]
\end{eqnarray}
In a similar way, the average work and heat for the third and fourth stroke are given by,
\begin{equation}
\label{4}
\la W_3\ra=\la H\ra_D-\la H\ra_C
=\left(\frac{\hbar\omega_1}{2}\,Q^\ast_2-\frac{\hbar\omega_2}{2}\right)\ct{\frac{\beta_2\hbar\omega_2}{2}}
\end{equation}
and
\begin{eqnarray}
\label{5}
 \la Q_4\ra&=&\la H\ra_A-\la H\ra_D \nonumber \\
&=&\frac{\hbar\omega_1}{2}\left[\ct{\frac{\beta_1\hbar\omega_1}{2}}-Q^\ast_2\,\ct{\frac{\beta_2\hbar\omega_2}{2}}\right].
\end{eqnarray}
For a heat engine, heat is absorbed from the hot reservoir, $\la Q_2\ra\geq0$, and flows into the cold reservoir, $\la Q_4\ra\leq0$. As a result, the two conditions have to be satisfied:
\be
Q^\ast_1\leq\frac{\ct{\beta_2\,\hbar\omega_2/2}}{\ct{\beta_1\,\hbar\omega_1/2}} \quad Q^\ast_2\geq\frac{\ct{\beta_1\,\hbar\omega_1/2}}{\ct{\beta_2\,\hbar\omega_2/2}}.
\ee
The efficiency of this quantum engine, defined as the ratio of the total work per cycle and the heat received from the hot reservoir, then follows as
\begin{eqnarray}
\label{6}
 \eta &=& -\frac{\la W_1\ra+\la W_3\ra}{\la Q_1\ra}\nonumber\\
&=&1-\frac{\omega_1}{\omega_2}\,\frac{\ct{\beta_1\,\hbar\omega_1/2} - Q^\ast_2\,\ct{\beta_2\,\hbar\omega_2/2}}{Q^\ast_1\,\ct{\beta_1\,\hbar\omega_1/2}-\ct{\beta_2\,\hbar\omega_2/2}}.\label{eff_exp}
\end{eqnarray}
The above exact expression is valid for arbitrary temperatures and frequency modulations, and allows for a detailed investigation of the performance of the  engine.

\paragraph{Efficiency at maximum power.}
Two essential characteristics of a heat engine are the power output, $P=-(\la W_1\ra+\la W_3\ra)/(\tau_1+\tau_2+\tau_3+\tau_4)$, and the efficiency at maximum power \cite{and84}.  Both can be evaluated analytically for the quantum Otto cycle with the help of Eq.~\eqref{6}. We shall separately consider the case of an adiabatic process, $Q^\ast_{1,2}=1$, and the case of a sudden switch of the frequencies for which  $Q^\ast_{1,2}= (\omega_1^2+\omega_2^2)/(2\omega_1\omega_2)$. Let us begin with the high temperature regime $\beta_i\hbar\omega_j \ll1, (i,j=1,2)$. The total work produced by the heat engine  for a quasistatic frequency modulation is given by
\be
\la W_1\ra +\la W_3\ra = \frac{1}{\beta_1}\left(\frac{\omega_2}{\omega_1}-1\right)+\frac{1}{\beta_2}\left( \frac{\omega_1}{\omega_2}-1\right).
\ee
Assuming that the initial frequency $\omega_1$ is fixed and by optimizing with respect to the second frequency $\omega_2$, we find that the power is maximum when $\omega_2/\omega_1= \sqrt{\beta_1/\beta_2}$. As a consequence, the efficiency at maximum power is
\be
\label{9}
\eta_\text{ad}= 1-\sqrt{\beta_2/\beta_1},
\ee
which corresponds to the Curzon-Ahlborn efficiency \cite{cur75}. Conversely, for a sudden switch, the total work is
\be
\la W_1\ra +\la W_3\ra = \frac{1}{2\beta_1}\left[\left(\frac{\omega_2}{\omega_1}\right)^2-1\right]+\frac{1}{2\beta_2}\left[\left( \frac{\omega_1}{\omega_2}\right)^2-1\right].
\ee
By optimizing again with respect to $\omega_2$, we find the power to be maximized when the condition $\omega_2/\omega_1= \sqrt[4]{\beta_1/\beta_2}$ is satisfied. The corresponding efficiency reads \cite{rez06},
\be
\label{12}
\eta_\text{ss}= \frac{1-\sqrt{\beta_2/\beta_1}}{2+\sqrt{\beta_2/\beta_1}}.
\ee
Equations \eqref{9} and \eqref{12} show that maximum efficiency (of either $1$ or $1/2$) can be attained when $\beta_1\rightarrow \infty$. In this low-temperature limit, quantum effects are crucial. Repeating the above {optimization} analysis in the regime $\beta_1\hbar\omega_1\gg1$, we find the efficiency at maximum power 
\be
\eta^\text{q}_\text{ad} = 1 - \sqrt{\hbar\omega_1\beta_2/2}
\ee
for an adiabatic process when  $\omega_2 = \sqrt{2\omega_1/\hbar \beta_2}$, and
\be
\eta^\text{q}_\text{ss} = \frac{1 - \sqrt{\hbar\omega_1\beta_2/2}}{2 + \sqrt{\hbar\omega_1\beta_2/2}},
\ee
for a sudden frequency switch when $\omega_2 = \sqrt[4]{2\omega_1^3/\hbar\beta_2}$. The above expressions, {in which the classical thermal energy ${k_B}T_1$ is replaced by the ground state energy $\hbar\omega_1/2$ of the oscillator,} are the quantum extensions of the Curzon-Ahlborn and Rezek-Kosloff results \eqref{9} and \eqref{12}.

\begin{figure}
\includegraphics[width=\columnwidth]{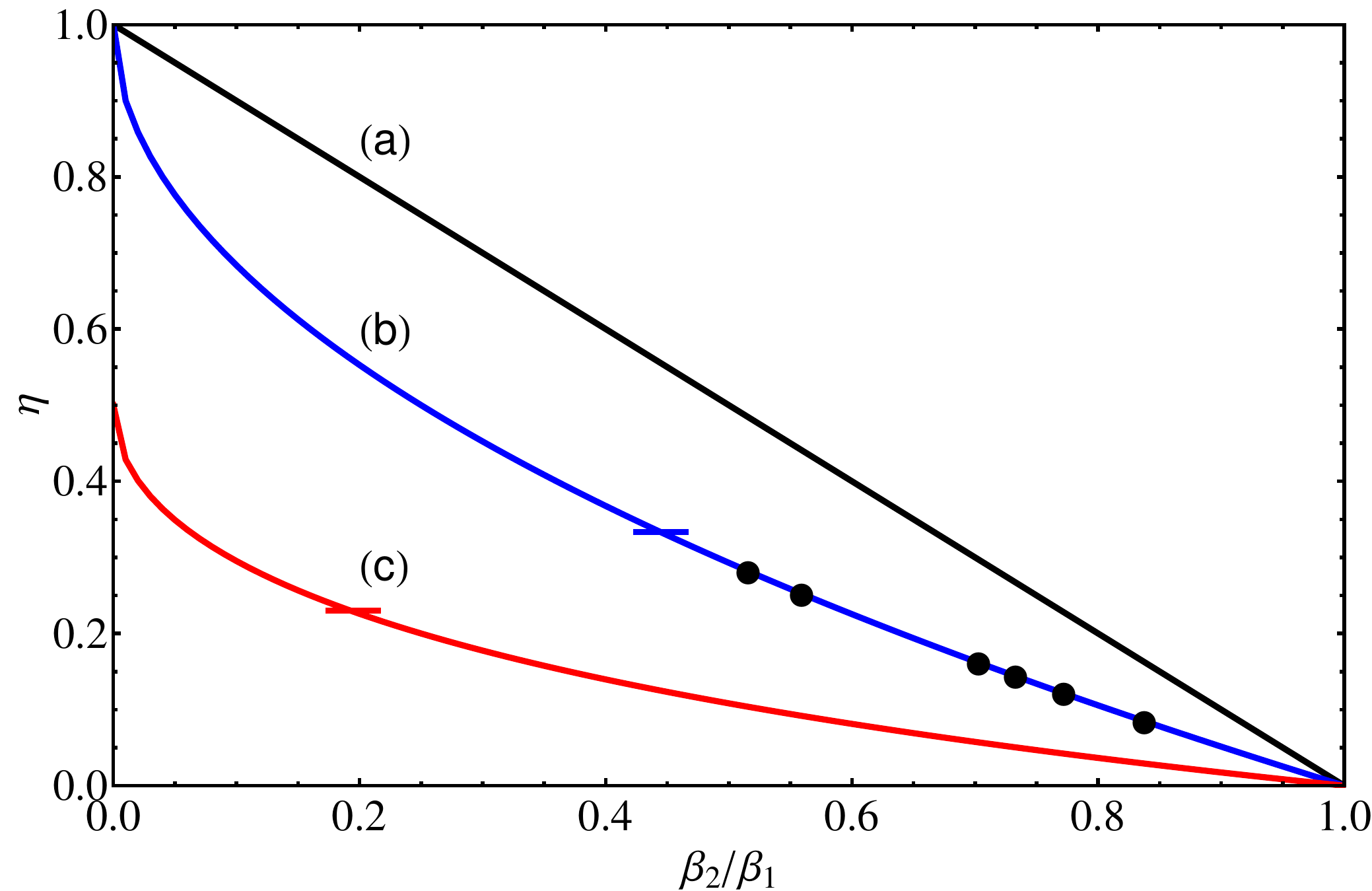}
\caption{(Color online) Efficiency at maximum power of the Otto engine as a function of the temperature ratio. (a) shows the Carnot efficiency with zero power. (b) corresponds to the theoretical adiabatic process, Eq. \eqref{9}, while (c) corresponds to the sudden frequency switch, Eq. \eqref{12}. The black points denote the results of the numerical simulations and demonstrate that the engine runs at maximum efficiency at maximum power. { The horizontal bars indicate  estimated upper bounds at $\omega_2= 1.5\, \omega_1$}.
\label{theory}}
\end{figure}

\paragraph{Proposed realization in a Paul trap.}
Such single ion heat engine is composed of one trapped ion in a modified linear Paul trap, as sketched in Fig.~\ref{cycle}(c).
The trapped ion is initially prepared in a thermal state at low temperature by laser cooling to the Doppler limit  in all spatial directions.
{The engine is driven} by alternatingly coupling the ion to two reservoirs that heat and cool the thermal state of the ion in the radial direction through scattering forces. {These two baths are realized} by blue and red detuned laser beams on a cycling transition of the trapped ion, irradiated in the radial plane ($x,y$). In a first step, {the coupling to the heat reservoirs is switched} on and off externally.
The geometry of a trap design with tapered radio-frequency (rf) electrodes leads to a pseudopotential of the form \cite{blinov04},
\be
V_p(x,y,z)=\frac{m}{2} \frac{  (\omega^2_{0x} \, x^2 + \omega^2_{0y} \, y^2)r_0^4}{(r_0 + z\,\tan\,\theta)^4}+\frac{m}{2} \omega_{0z}^2\,z^2 ,
\ee
where $\theta$ is the angle between the electrodes and the trap axis $z$, and $r_0$ the radial distance of the ion to the electrodes, as shown in Fig.~\ref{cycle}(c). This potential results in radial trap frequencies that depend on the axial position~$z$, and in an axial force that depends on the radial displacement. The coupling between harmonic axial and radial modes is of the {generic} form,
$H=\sum_{i\in\left\{x,y,z\right\}} \hbar \, \omega_{0i}\, ( 1/2 + a_i^\dagger \, a_i  ) - C\, \hat{z} \cdot (\omega_{0x}^2 \hat{x}^2 + \omega_{0y}^2 \hat{y}^2)$, valid for small $z$, 
where ${C=2\,m\tan\theta/r_0}$ denotes the coupling constant between the oscillator modes, and $\omega_{0i}$ are the trap frequencies at the center of the trap.
A change in temperature $T$ of the radial thermal state of the ion, and { thus} of the width of its spatial distribution, leads to a { modification} in the axial component of the repelling force { which changes} the point of equilibrium $z_0(T)$.
Heating and cooling the radial  state hence moves the ion back and forth along the trap axis, as sketched in Fig.~\ref{cycle}(b), resulting in the closed Otto cycle shown in Fig.~\ref{cycle}(a). This thermally induced axial movement corresponds to the mechanically usable movement of a piston of a classical engine, while the radial mode  corresponds to the gas in the cylinder.

The energy gained by running the engine cyclically in the radial direction can be stored in the axial degree of freedom. Indeed, the axial translation on the nanometer scale can be enhanced by coupling the ion to the laser reservoirs resonantly at the axial trap frequency. 
In such a way, the cyclic  temperature change of the radial thermal state is directly converted into an increasing coherent axial oscillation.
In principle, the excited axial oscillation is only limited by the trap geometry. {The cycle time of the engine is given by the axial oscillation period, and is of the order of $10\,\mu$s, while the time needed to change the temperature  is below $1\,\mu$s. In order to reach a steady state of the heat engine,  {a  red-detuned low intensity dissipation laser is applied} in axial direction that damps the coherent movement depending on the velocity of the ion.}
The stored energy in the axial motion may be transferred to other oscillator systems, e.g. separately trapped ions \cite{harlander10} or nanomechanical oscillators \cite{Tian04}, and thus extracted from the engine as usable work. When driven as a heat pump, cooling of such systems should be possible.

The single ion  engine  offers complete control of all parameters over a wide range. The temperature of the two heat reservoirs, as well as  the applied dissipation, can  be adjusted by tuning the laser frequencies and intensities. On the other hand, the oscillation amplitude and frequency of the ion can be changed by virtue of the trap parameters. This unique flexibility of the device can be exploited to satisfy the conditions for maximum efficiency at maximum power derived previously.

\paragraph{Monte Carlo simulations.}
We have performed extensive semiclassical simulations of the  engine using Monte-Carlo and partitioned Runge-Kutta integrators, as described in \cite{singer2010}. {We reach a dynamic confinement of the ion in the trap volume through  an oscillating parabolic and tapered saddle potential of the form}
\be
V_\text{rf} \propto \frac{U_0 \sin(\omega_\text{rf} t)}{(r_0+z \, \tan \theta)^2} \cdot (x^2-y^2),
\ee
with a { trap drive} frequency $\omega_\text{rf}=60\,$MHz, and the trap geometry described in Fig.~\ref{cycle}(c). The thermal state of the ion is generated by a Boltzmann distributed ensemble over several thousand classical trajectories \cite{casdorffBlatt88}: the initial parameters for each ion are choosen randomly, and the desired thermal probability distribution is reached through the ion-light interaction and the corresponding stochastic spontaneous emission of photons \cite{srednicki94}.
\begin{figure}
\includegraphics[width=\columnwidth]{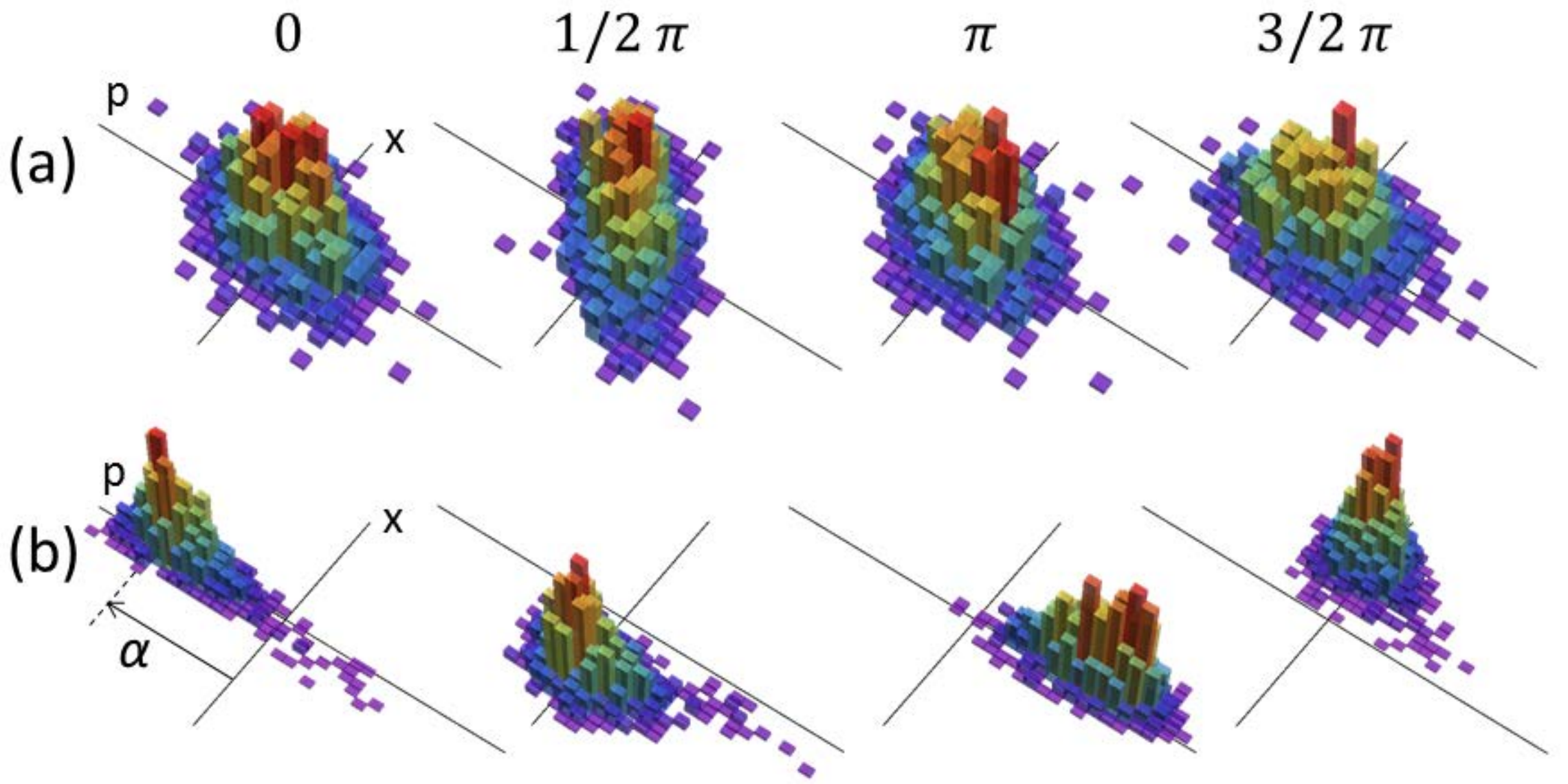}
\caption{(Color online) Motional-state phase space distribution during an engine cycle $ABCD$: (a) radial mode changing from thermal to nonthermal states. (b) In axial direction, a coherent oscillation with {typical displacement of $\alpha= 10^3$ wave packets} displays the functional performance of the engine.
\label{phasespace}}
\end{figure}
{The switching of the detuned heating and cooling lasers is adjusted} to the axial trap frequency $\omega_z$. Each laser is coupled to the ion for $20\%$ of an axial trap period. The ensemble of driven oscillators is thus excited coherently in axial direction such that heating/cooling takes place at the turning points of the trajectory, independently of the random initial conditions. In phase space, the ensemble average of the axial mode performs a phase synchronized oscillation (see  Fig.~\ref{phasespace}), while the radial mode changes between thermal and nonthermal states.

Radial temperatures in the range of $20$ to $200\,$mK were  achieved, corresponding to $0.1<\beta_2/\beta_1<1$ and respective  \mbox{radial phonons numbers of  about $400$ and $4000$}. For a realistic  maximal axial amplitude of about $1\,$mm, the  relative variation of the radial frequency at $6.0\,$MHz is about $50\%$.
By properly adjusting the parameters to satisfy the optimality condition, {our simulations show} that this Otto engine {has the ability} to run at maximum efficiency at maximum power in the interval $0.5<\beta_2/\beta_1<1$  (see Fig.~\ref{theory}). The  maximum efficiency, determined through Eq.~\eqref{eff_exp} with values taken from simulated cycles as shown in Fig.~\ref{cycle}(a),  is about $30\,\%$. This maximum efficiency is significantly larger than those obtained to date \cite{bechinger11, steeneken11}. The expected power of the engine is of the order  of $P\simeq  10^{-20}$J/s.
The case of a sudden switch can be  realized by exchanging the values of radial and  axial frequencies. Contrary to the adiabatic limit, the optimality condition can be achieved down to $\beta_2/\beta_1 \simeq 0.2$.
However, the maximum efficiency {reduces to} $23\,\%$.

While so far  {the laser reservoirs have, for simplicity, been switched}  on and off externally, another feature of this Otto engine is the ability to operate in a completely self-driven manner. To this end, the foci of the heating and cooling lasers {can be} separated spatially on the trap axis by e.g. $200\,\mu$m, so that the ion is coupled to the heat baths at the turning points of its axial trajectory. No {active} switching is required and  the axial motion of the ion is self-amplifying. The ion needs to be driven only in the initial phase of the axial motion to reach a threshold amplitude.
We note that at a radial frequency of $6.0\,$MHz, the thermal energy of the oscillator, Eq.~(1a), starts to deviate appreciably from its classical value below $200\,\mu$K, which could be reached with $^{40}$Ca-ions, if we assume a two-level approximation and the Doppler cooling limit 
$T_D= \hbar \Gamma / (2 { k_B})$, where $\Gamma$ denotes the linewidth of the dipole transition.  A single ion engine has thus the potential to enter the quantum regime and {become a  tool to  study effects of quantum coherence and correlations on the  efficiency.  Application of optimal control techniques \cite{schmitt2011} would further allow for nonclassical bath engineering.} {The investigation of heating and cooling on the simple and
fundamental single ion mode interactions may serve for
prototyping coolers also in systems,  which share similar properties. For the specific example of micromechanical oscillators, mode coupling has been described and realized in several experiments \cite{kar09,fau12}.}

\paragraph{Conclusion.}
We have put forward a realistic proposal for a tunable nanoengine based on a single ion in a tapered linear Paul trap coupled to engineered laser reservoirs. The operation in the Otto cycle would result in coherent ion motion. Combining  analytical and numerical analysis, we have studied the performance of the engine and showed that it can achieve maximum efficiency at maximum power in a wide range of temperatures. {We expect a stimulating impact on the development on nano- and micromechanical oscillators which are of fundamental importance for future sensor technologies.
}

\paragraph{Acknowledgments.} {We thank U. Poschinger for carefully  reading the manuscript}. This work was supported by the Emmy Noether Program of the DFG (contract no LU1382/1-1), the cluster of excellence Nanosystems Initiative Munich (NIM), the Alexander-von-Humboldt Foundation, the Volkswagen-Stiftung, the DFG-Forschergruppe (FOR 1493) and the EU-project DIAMANT (FP7-ICT).


\begin{thebibliography}{99}
\bibitem{cer09} G. Cerefolini, {\it Nanoscale Devices}, (Springer, Berlin, 2009).
\bibitem{cen01} Y.A. Cengel and M.A. Boles, {\it Thermodynamics. An Engineering Approach}, (McGraw-Hill, New York, 2001).
\bibitem{kay07} E.R. Kay, D.A. Leigh, and F. Zerbetto, Angew. Chem. Int. Ed. {\bf 46}, 72 (2007).
\bibitem{mar09} P. H\"anggi and F. Marchesoni, Rev. Mod. Phys. {\bf 81}, 387 (2009).
\bibitem{sco59} H.E.D. Scovil and E.O. Schulz-DuBois, Phys. Rev. Lett. \textbf{2}, 262 (1959).
\bibitem{ali79} R. Alicki, J. Phys. A {\bf 12}, L103 (1979).
\bibitem{kos84} R. Kosloff, J. Chem. Phys. \textbf{80}, 1625 (1984).
\bibitem{gev92} E. Geva and R. Kosloff, J. Chem. Phys. \textbf{96}, 3054 (1992);  {\bf 97}, 4398 (1992).
\bibitem{scu02} M.O. Scully, Phys. Rev. Lett. {\bf  88}, 050602 (2002).
\bibitem{hum02} T.E. Humphrey \etal, Phys. Rev. Lett. {\bf 89}, 116801 (2002).
\bibitem{scu03} M.O. Scully \etal, Science {\bf 299}, 862 (2003).
\bibitem{kie04} T.D. Kieu, \PRL \textbf{93}, 140403 (2004).
\bibitem{dil09} R. Dillenschneider and E. Lutz, EPL \textbf{88},  50003 (2009). 
\bibitem{gem09} J. Gemmer, M. Michel, and G. Mahler, \textit{Quantum Thermodynamics} (Springer, Berlin, 2009).
\bibitem{hug02} T. Hugel \etal, Science {\bf 296}, 1103 (2002).
\bibitem{jac03} S.A. Jacobson and  A.H. Epstein, in Proc. Int. Symp. Micro-Mechanical Engineering, pp. 513-520 (2003).
\bibitem{wha03} S. Whalen \etal, Sensors and Actuators {\bf 104}, 290 (2003).
\bibitem{steeneken11} P.G. Steeneken \etal, Nature Phys. \textbf{7}, 4 (2011).

\bibitem{bechinger11} V. Blickle and C. Bechinger, Nature Phys. {\bf 8}, 143 (2012).
\bibitem{lei03} D. Leibfried \etal, \RMP {\bf 75}, 281 (2003).
\bibitem{poyatos96} J.F. Poyatos, J.I. Cirac, and P. Zoller, \PRL {\bf 77}, 23 (1996).

\bibitem{blatt08} R. Blatt, and D. Wineland, Nature {\bf 453}, 1008 (2008).
\bibitem{monz11} T. Monz \etal, \PRL {\bf 106}, 130506 (2011).
\bibitem{hub08} G. Huber \etal, \PRL \textbf{101}, 70403 (2008).
\bibitem{lin03} B. Lin and J. Chen, \PRE \textbf{67}, 046105 (2003).
\bibitem{rez06} Y. Rezek and R. Kosloff, \NJP \textbf{8}, 83 (2006).

\bibitem{qua07} H.T. Quan \etal, \PRE \textbf{76}, 031105 (2007).
\bibitem{and84} B. Andresen, P. Salamon, and R.S. Berry,  Phys. Today \textbf{37}, 62 (1984).
\bibitem{hus53} K. Husimi, Prog. Theo. Phys. \textbf{9}, 381 (1953).
\bibitem{mahler05} F. Tonner and G. Mahler, \PRE \textbf{72}, 66118 (2005).
\bibitem{OConell10} A.D. O'Connell \etal, Nature \textbf{464}, 697 (2010).
\bibitem{lin10} Q. Lin \etal, Nature Phot. \textbf{4}, 236 (2010).
\bibitem{kar09} R.B. Karabalin \etal, Phys. Rev. B \textbf{79}, 165309 (2009).
\bibitem{fau12} T. Faust \etal, arXiv:1201.4083. 

\bibitem{def08} S. Deffner and E. Lutz, \PRE \textbf{77},  021128 (2008).
\bibitem{def10} S. Deffner, O. Abah, and E. Lutz, Chem. Phys. \textbf{375}, 200 (2010).
\bibitem{cur75} F.L. Curzon and B. Ahlborn, Am. J. Phys. \textbf{43}, 22 (1975).

\bibitem{harlander10} M. Harlander \etal, Nature {\bf 471}, 200 (2011).
\bibitem{blinov04} B.B. Blinov \etal, Quantum Inf. Process. \textbf{3}, 45 (2004).
\bibitem{Tian04} L. Tian and P. Zoller, \PRL \textbf{93}, 266403 (2004).
\bibitem{singer2010} K. Singer \etal, \RMP \textbf{82}, 3 (2010).
\bibitem{casdorffBlatt88} R. Casdorff and R. Blatt, Appl. Phys. B \textbf{45}, 3 (1988).
\bibitem{srednicki94} M. Srednicki, \PRE \textbf{50}, 2 (1994).
\bibitem{schmitt2011} M. Schmidt \etal, \PRL \textbf{107}, 130404 (2011).

\end{thebibliography}
\end{document}